\documentclass[pra,twocolumn,showpacs,floatfix]{revtex4}
\usepackage{graphicx}
\usepackage{amsmath}
\usepackage{amssymb}
\usepackage{epsf,latexsym}
\usepackage{times}
\usepackage{bm}

\usepackage{float}

\newcommand{\be}{\begin{equation}}
\newcommand{\ee}{\end{equation}}
\newcommand{\ba}{\begin{eqnarray}}
\newcommand{\ea}{\end{eqnarray}}

\newcommand{\bi}{\bibitem}

\renewcommand{\r}{({\bm r})}

\begin{document}
\title{Reverse engineering in many-body quantum physics: What many-body 
system corresponds to an effective single-particle equation?}
\author{J. P. Coe$^{1,2}$
}
\email{jpc503@york.ac.uk}
\author{K. Capelle$^{3,4}$
}
\author{I. D'Amico$^{1}$
}
\email{ida500@york.ac.uk}
\affiliation{
$^1$ Department of Physics, University of York, York YO10 5DD, United Kingdom.\\
$^2$ Department of Mathematics, University of York, York YO10 5DD, United Kingdom.\\
$^3$ Departamento de F\'{i}sica e Inform\'{a}tica, Instituto de F\'{i}sica de S\~{a}o Carlos, Universidade de S\~{a}o Paulo, Caixa Postal 369, 13560-970 S\~{a}o Carlos, SP, Brazil.\\
$^4$ Theoretische Physik, Freie Universit\"at Berlin, D-14195 Berlin,
Germany.}
\date{\today}

\begin{abstract}
The mapping, exact or approximate, of a many-body problem onto an effective 
single-body problem is one of the most widely used conceptual and computational
tools of physics. Here, we propose and investigate the inverse map of effective 
approximate single-particle equations onto the corresponding many-particle 
system.
This approach allows us to understand which interacting system a given 
single-particle approximation is actually describing, and how far this is 
from the original physical many-body system. 
We illustrate the resulting reverse engineering process by means of the
Kohn-Sham equations of density-functional theory. In this application, our
procedure sheds light on the non-locality of the density-potential mapping
of DFT, and on the self-interaction error inherent in approximate density 
functionals.
\end{abstract}


\maketitle

One of the most widely used and successful approaches to many-particle physics 
is to map the many-body problem onto an effective single-body problem.
Innumerable concepts and methods of theoretical physics derive from this
general idea. The mean-field approximation, one of the most widely used
approximation schemes in all fields of physics, is of this type, as well
as the (in principle exact) mapping onto Kohn-Sham (KS) equations, used
in density-functional theory (DFT). Here we introduce the concept of
{\em reverse engineering of single-particle equations} \cite{footnote1}, 
as a tool for discovering for which many-body system a given single-particle 
approximation becomes exact. The idea is developed below in the framework 
of DFT, but the concept is completely general, and can be extended to other 
single-particle methods.

DFT \cite{KOHN} in the Kohn-Sham formulation \cite{KS} is a 
tool for calculating properties of many-body systems by means of a one-to-one 
mapping between the interacting system and a fictitious non-interacting one, 
with the same ground-state density $n\r$. All ground-state properties can in 
principle be expressed as functionals of this density. The success of DFT 
depends on the quality of the approximation to the exchange-correlation (xc)
energy functional $E_{\text{xc}}[n]$, which enters the KS equations through 
its functional derivative, the xc potential $v_{\text{xc}}\r$. $E_{\text{xc}}$ 
stems from the interactions in the original many-body system, and its 
functional form for Coulomb interacting systems, such as electrons in atoms,
molecules, nanostructures and solids, is unknown.
In order to construct viable approximations, a great deal of work has therefore been devoted to the derivation of exact properties of the exchange-correlation 
functional and potential \cite{LEVYPERDEW94,PERDEWETAL07}. The performance of a 
functional is judged by how close the density, and the observables calculated 
from it, are to that of the many-body system under study. 

The idea of reverse engineering suggests a different mode of analysis, namely 
to ask: for what system does a given approximation become exact? This question 
can be interpreted in two ways, one very common, the other being proposed here.
To exemplify both, consider the local-density approximation (LDA).
By construction, the LDA becomes exact for uniform densities. But in 
practice we rarely apply density functionals to uniform systems. Instead, we 
apply DFT to realistic inhomogeneous many-body systems, for which the LDA 
density is ${\bm r}$-dependent and approximate. The question we ask is: 
for which alternative many-body system is this approximate density the exact 
ground-state density? In this context, we call this alternative many-body 
system the interacting-LDA (i-LDA) system \cite{Coe}. Reverse engineering of 
DFT refers to the process of constructing this alternative many-body system. 

The aim of this construction is not to simulate large systems, for which 
nothing is gained by mapping a many-body system on another, equally complex, 
one. Rather, it shifts attention from the density predicted by the LDA to
the external potential predicted by the LDA (i.e., the i-LDA potential),
allowing us to investigate the LDA (or any other single-particle approximation 
to the many-particle Hamiltonian) in a novel and particularly detailed way.
For example, we can now ask: how close is the 
corresponding i-system external potential to the true external potential? 
{\it Which artificial features is the chosen approximation building into the 
system?} As in other parts of science, reverse engineering enables one to
understand the functionality and structure of the engineered device
on a different level, opening up new pathways for improvement.

To illustrate the basic idea, we start from the LDA densities of the 
helium atom and of Hooke's atom, and invert the many-body Schr\"odinger 
equation to construct that external potential for which the LDA densities are 
exact ground-state densities. Comparison to the true external potentials
of the helium atom ($\sim 1/r$) and Hooke's atom ($\sim r^2$) reveals the
errors inherent in the approximate density and functional.  By calculating explicitly the external potential for the i-LDA system, the proposed method provides the exact i-LDA system Hamiltonian. This implies that all properties of the i-LDA system (which themselves depend on the accuracy of the LDA) can, in principle, be directly calculated and compared to the ones of the exact many-body system. 

Studying relatively simple systems, such as helium or Hooke's atoms, to understand approximations has proved fruitful in the 
past \cite{BAERENDS01,1DHELIUMTDFT,LEIN}.  Earlier inversion
schemes in DFT were used to find the $v_{\text{xc}}\r$ that reproduces a given 
exact density, by inverting the single-particle Kohn-Sham equations \cite{INVERSIONREFS}. This approach 
provides information on the exact $v_{\text{xc}}$. The reverse engineering 
procedure, by contrast, aims at reproducing a given approximate density by 
inverting the {\em many-body} Schr\"odinger equation, providing information 
on the approximate $v_{\text{xc}}$. Inversion of the many-body equation is 
a much harder task, which 
up to now had only been attempted for a one-dimensional model system within 
the adiabatic approximation to time-dependent DFT \cite{1DHELIUMTDFT}. To 
achieve this, we developed inversion schemes that are substantially more 
accurate then previous ones (see discussion below). 

We consider the N-body Hamiltonian 
$H=\hat{T}+\hat{V}_{\text{ee}}+\hat{V}_{\text{ext}}$,
where $\hat{T}$ is the kinetic energy operator, $\hat{V}_{ee}$ is the 
electron-electron interaction and 
$\hat{V}_{\text{ext}}=\sum_{i=1}^{N}v_{\text{ext}}(\bm{r}_{i})$,
is the external potential. By multiplying the Schr\"odinger equation 
$\hat{H}\Psi=E\Psi$ from the left by $\Psi^{*}$ and integrating over all 
but one of the coordinates we obtain
\begin{eqnarray}
\nonumber\int\Psi^{*}(\bm{r_{1}}\ldots\bm{r_{N}})\left(\hat{T}+\hat{V}_{\text{ee}}+\hat{V}_{\text{ext}}\right)\\
\Psi(\bm{r_{1}}\ldots\bm{r_{N}})d^{3}r_{2}\ldots d^{3}r_{N}=E \frac{n(\bm{r_{1}})}{N}.
\end{eqnarray}
From this we find, for the term concerning the external potential,
\begin{eqnarray}
\nonumber & & \int \Psi^{*} \sum_{i=1}^{N}v_{\text{ext}}(\bm{r_{i}})\Psi d^{3}r_{2}\ldots d^{3}r_{N} 
=v_{\text{ext}}(\bm{r}_{1})\frac{n(\bm{r_{1}})}{N}+ \\
& & (N-1)\int \Psi^{*} v_{\text{ext}}\bm{(r_{2}})\Psi d^{3}r_{2}\ldots d^{3}r_{N} 
\end{eqnarray}
\begin{align}
=v_{\text{ext}}(\bm{r_{1}})\frac{n(\bm{r_{1}})}{N}+ 
\frac{2}{N}\int \gamma(\bm{r_{1}},\bm{r_{2}};\bm{r_{1}},\bm{r_{2}})v_{\text{ext}}(\bm{r_{2}})d^{3}r_{2}, 
\end{align}
where $\gamma(\bm{r_{1}},\bm{r_{2}};\bm{r_{1}},\bm{r_{2}})=\frac{N(N-1)}{2}\int \Psi^{*} \Psi d^{3}r_{3}\ldots d^{3}r_{N}$.

We combine these results to obtain an iterative relation for the external 
potential $v_{\text{ext}}(\bm{r_{1}})$ that reproduces the target density 
$n^{\text{target}}(\bm{r_{1}})$,
\begin{eqnarray}
\nonumber v^{i+1}_{\text{ext}}(\bm{r_{1}})=\frac{1}{n_i(\bm{r_{1}})}
\big[ E_i n^{\text{target}}(\bm{r_{1}})  \\  \nonumber
-N\int \Psi_i^{*}(\hat{T}+\hat{V}_{\text{ee}} )\Psi_i d^{3}r_{2}
\ldots d^{3}r_{N}  \\ 
-2\int \gamma_i(\bm{r_{1}},\bm{r_{2}};\bm{r_{1}},\bm{r_{2}})
v^{i}_{\text{ext}}(\bm{r_{2}})d^{3}r_{2} \big] \label{eqn:vext1}. 
\end{eqnarray}
To avoid having to calculate the integrals, we use the identity 
\begin{eqnarray}
\nonumber n_i(\bm{r_{1}})v^i_{\text{ext}}(\bm{r_{i}})-E_i n_i(\bm{r_{1}})= \\ \nonumber 
-N\int \Psi_i^{*}(\hat{T}+\hat{V}_{\text{ee}} )\Psi_i d^{3}r_{2}\ldots d^{3}r_{N}\\
-2\int \gamma_i(\bm{r_{1}},\bm{r_{2}};\bm{r_{1}},\bm{r_{2}})v^i_{\text{ext}}(\bm{r_{2}})d^{3}r_{2}, \label{eqn:videntity} 
\end{eqnarray}
and obtain the simple iterative relation
\begin{equation}
v^{i+1}_{\text{ext}}(\bm{r_{1}})=\frac{1}{n_i(\bm{r_{1}})}E_i[n^{\text{target}}(\bm{r_{1}})-n_i(\bm{r_{1}})]+v^{i}_{\text{ext}}(\bm{r_{1}}),
\label{eq:heliumscheme}
\end{equation}
which is to be iterated with the Schr\"odinger equation
$(\hat{T}+\hat{V}_{\text{ee}}+\hat{V}_{\text{ext}}^i)\Psi_i=E_i\Psi_i$.
At convergence, $n_i(\bm {r_{1}})\equiv n^{\text{target}}(\bm {r_{1}})$, and 
$v_{\text{ext}}^{i+1}(\bm {r_{1}})\equiv v_{\text{ext}}^i(\bm {r_{1}})$ is the 
external potential that reproduces this density via the many-body Schr\"odinger
equation.

We note that if $n_{i}(\bm{r_{1}})$ is larger (smaller) than $n^{\text{target}}(\bm{r_{1}})$ then the potential must increase (decrease) at this point to bring $n_{i+1}(\bm{r_{1}})$ closer to $n^{\text{target}}(\bm{r_{1}})$. Therefore, iteration of Eq.~(\ref{eq:heliumscheme}) is expected to converge if $E_{i}<0$. If $E_{i}>0$, we replace 
Eq.~(\ref{eq:heliumscheme}) by
\begin{eqnarray}
v^{i+1}_{\text{ext}}(\bm{r_{1}})=\frac{1}{n_i(\bm{r_{1}})}E_i[n_i(\bm{r_{1}})-n^{\text{target}}(\bm{r_{1}})]
+v^{i}_{\text{ext}}(\bm{r_{1}}).
\label{eq:hookscheme}
\end{eqnarray}
Scheme (\ref{eq:heliumscheme}) converges relatively easily for the helium 
atom (where $E<0$) and scheme (\ref{eq:hookscheme}) for Hooke's atom (where $E>0$). We aid 
convergence in both cases by mixing $v^{i+1}_{\text{ext}}$ with $80\%$ 
of $v_{\text{ext}}^i$, and iterate until the relative error
$\int d^3r\, |n_i(\bm{r})-n^{\text{target}}(\bm{r})|/\int d^3r\,n^{\text{target}}(\bm{r})$
has reached a desired level.

At convergence, we obtain that external potential whose many-body ground
state has the same density as was predicted by the approximate density 
functional for the true external potential. As this interacting system ground-state reproduces the LDA density then by the Hohenberg-Kohn theorem \cite{HK} this interacting system must be the {\it exact} interacting system corresponding to the LDA. We check this property by independently solving the many-body Schr\"odinger equation with the converged 
alternative external potential, using a larger basis set than that used 
in the iterations. This additional consistency procedure allows our inversion scheme to be precise even in regions of space where the density is just 
$10^{-7}$ a.u. (helium) and $10^{-12}$ a.u. (Hooke's atom), i.e. orders of 
magnitudes smaller than previous inversion schemes, which, according to 
Ref.~\cite{1DHELIUMTDFT} attain an accuracy of up to $10^{-2}$ a.u.

We now illustrate both inversion schemes, and the additional consistency check, 
by applying our procedure to the helium atom and Hooke's atom, generating the approximate (target) density from the LDA. For two electrons in a spherically symmetric potential the ground state can only be a function of the distance of each electron from the origin and the angle between the electron vectors.  Hence we employ the basis
$\phi_{ijl}=R_{i}(r_{1})R_{j}(r_{2})\sqrt{2l+1}P_{l}(\cos(\theta))/(4\pi)$.
For Hooke's atom $R_{i}(r)=Q_{i}(r)e^{-\alpha r^{2}}$ is a harmonic 
oscillator-like wave-function while for the helium atom it is a hydrogen-like 
wave-function $R_{i}(r)=Q_{i}(r)e^{-\alpha r}$. The $Q_{i}(r)$ are polynomials 
of degree $i$ created via the Gram-Schmidt procedure such that the $R_{i}$ are 
orthonormal.

Hooke's atom is an interacting system of two electrons in the harmonic confining potential $v_{\text{ext}}=\omega r^{2}/2$. We use the LDA in the parametrization of Perdew and Wang \cite{PERDEW92} for $v_{\text{xc}}$, and solve the KS equations for $\hbar\omega=0.0365$ Hartree. For comparison we also calculate the exact density from the exact interacting wave function using the method of Taut \cite{TAUT}. We then apply Eq.~(\ref{eq:hookscheme}) to find the external potential of the interacting system that reproduces the LDA density, {\em i.e.} the i-LDA system \cite{Coe}.

In practice, an external potential found with a basis of e.g. $6^{3}$ 
functions may not reproduce the LDA density when we counter-check the scheme by solving the many-body Schr\"{o}dinger equation of the i-LDA system using a larger basis of size e.g.  $7^{3}$.  Hence, we repeat the procedure with increasing basis size until we find an external potential that reproduces the LDA density even with a larger basis. 

\begin{figure}[ht]\centering
  \includegraphics[width=.38\textwidth]{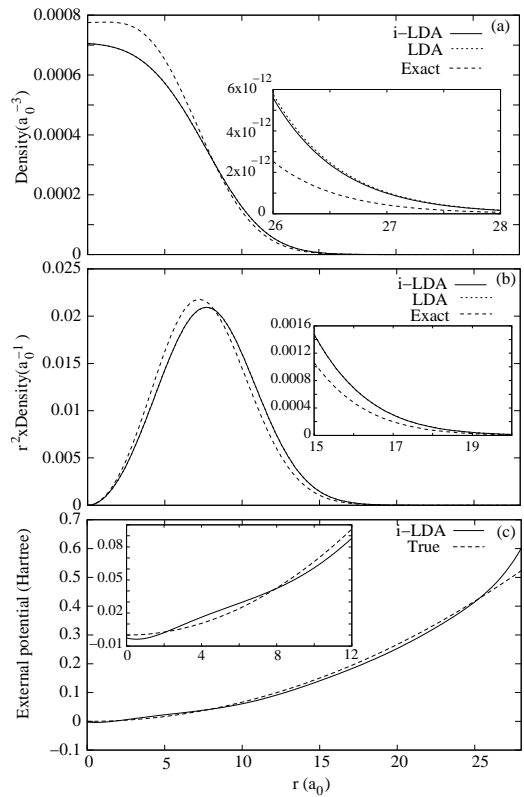}
  \caption{Hooke's atom: $\hbar \omega=0.0365$ Hartree, basis size: $8^3$ functions (counter-checked with a $9^3$ basis set, see text). (a) Comparison of LDA, exact and i-LDA densities. Inset: Zoom of the LDA and i-LDA densities at large $r$. (b) Comparison of LDA, exact and the i-LDA radial probability densities. Inset: Zoom of the tail of the density. (c) Comparison of the true external potential and the i-LDA external potential. Inset: Zoom of the potentials close to the origin. }\label{fig:NEWn=4fusion}
\end{figure}

In Fig.~\ref{fig:NEWn=4fusion} (a) the exact and LDA densities are plotted. The LDA and i-LDA densities are indistinguishable, but the difference 
between the LDA and exact densities is large, and consequently the potentials 
of the true and the i-LDA system are rather different. 
Fig.~\ref{fig:NEWn=4fusion}(c) shows a substantial difference between 
the i-LDA potential and the true external potential at large $r$, where the 
i-LDA potential grows  more rapidly than the true potential. The LDA 
density is very different from the true density in this region (Fig.~\ref{fig:NEWn=4fusion} (a) inset), with a relative error $(n(r)-n^{\text{LDA}}(r))/n^{\text{LDA}}\sim 60\%$ for $r=28$ \cite{note1}. By contrast, 
near the origin the i-LDA potential is slightly weaker than the true one. 
In between the two limiting regions, the i-LDA potential and the true external 
potential cross various times. These crossings manifest themselves in a complex
way in the behavior of the corresponding densities: Fig.~\ref{fig:NEWn=4fusion} shows that a crossing in the potentials around $r=8$ (Fig.~\ref{fig:NEWn=4fusion} (c) inset) corresponds to a crossing in the densities (Fig.~\ref{fig:NEWn=4fusion} (b)), whereas the crossings at $r\sim 2$ and $r\sim 25$ are not accompanied by a crossing in the densities. 

Next we consider the helium atom, for which we use the iterative scheme of Eq.~(\ref{eq:heliumscheme}) to calculate the i-LDA external potential. Here a basis of $7^{3}$ functions is required to find the external potential that satisfies our consistency check, i.e. reproduces the density when solving the Schr\"odinger equation with a larger basis ($8^{3}$ functions).

\begin{figure}[ht]\centering
  \includegraphics[width=.38\textwidth]{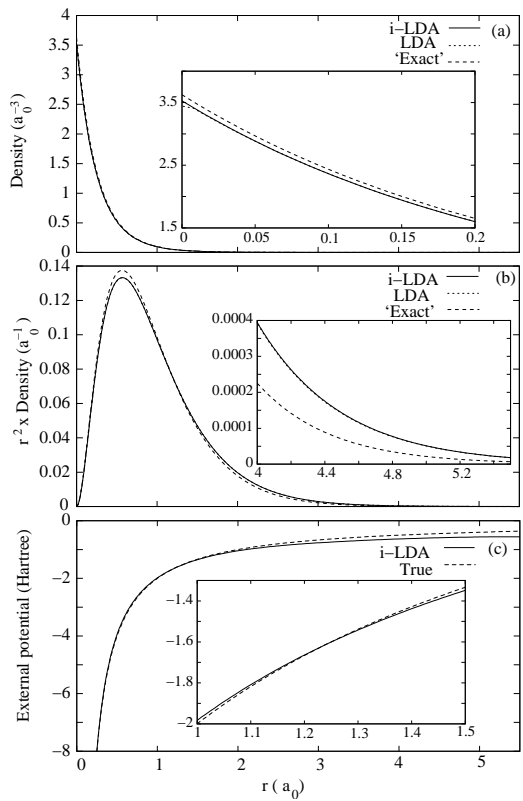}
  \caption{The helium atom: (a) Comparison of LDA, exact and i-LDA densities.  Inset: zoom of the densities close to the origin. 
(b) Comparison of LDA, exact and the i-LDA radial probability densities. 
Inset: Zoom of the tail of the density.  (c) Comparison of the true external 
potential and the i-LDA external potential.}\label{fig:NEWHeliumfusion}
\end{figure}

As Fig.~\ref{fig:NEWHeliumfusion} shows, the LDA reproduces relatively well the exact radial probability density, but an underestimate at small $r$ and overestimate at large $r$ is evident. A closer view of large $r$ (inset of Fig.~\ref{fig:NEWHeliumfusion}(b)) shows that there is a larger discrepancy ($49\%$ on the range $4 \leq r/a_{0} \leq 5.5$ compared to $3.68\%$ overall) between the LDA density and the exact density here.  However we still find that the i-LDA system density is almost indistinguishable from the LDA density on this scale ($0.78\%$ error) and its overall error at $0.037\%$ is comparable to that of $0.0063\%$ which we achieved for Hooke's atom. In Fig.~\ref{fig:NEWHeliumfusion}(c) we again note that the external potential of the i-LDA system is substantially different from the true one for large values of $r$, where it is weaker than the true potential, causing the LDA density to be spread out slightly more than the exact one.  We also observe a crossing of both  potentials and densities for $r\sim 1.2$ (Fig.~\ref{fig:NEWHeliumfusion}(b)) where the radial probability density is high.

Both Hooke's and the helium atom thus exhibit a relationship between crossings in 
densities and crossings in potentials, but this relation is local only in 
regions where the radial probability density is high.  The systems considered here have spherical symmetry, hence the 
radial probability density $4\pi n(r) r^{2}$ indicates the probability of finding a particle at 
distance $r$ from the origin. Thus, a single marked peak of the radial probability density 
at $r$ indicates a high probability of finding a particle at that distance and 
a very low probability of finding the particle elsewhere. In this sense, the 
system is ``almost classical'' in the region of the peak. The charge density itself does
not have this property.
The regions of high radial probability density can be interpreted as almost classical, so that a local relationship 
between the `particle position' and the external potential can be expected.

By contrast, crossings 
in the potentials in regions with low values of the radial probability density 
do not necessarily correspond to crossings in the densities.
These non-local regions are the
ones in which the quantum nature of the system is more apparent.  The identification of such regions is particularly interesting in view of recent investigations of semiclassical approximations to DFT \cite{ELLIOT08}. 
The LDA density can even be higher than the exact density in regions where the i-LDA external
potential is more repulsive than the true external potential. This dramatically
highlights the non-local nature of the density-potential mapping in DFT

Apart from the non-locality of the density-potential mapping, the other main 
feature common to both systems is the substantial difference between the i-LDA 
and exact external potential at large $r$. For Hooke's atom in particular, the 
i-LDA potential diverges to infinity even faster than the exact 
potential \cite{note2}. We explain
this as a consequence of the erroneous asymptotic decay of the LDA 
exchange-correlation potential (exponential rather than $1/r$), which itself 
is a consequence of the single-electron self-interaction error inherent in 
the LDA. Due to this error, the LDA $xc$ potential is too weak in the 
asymptotic region, so that the i-LDA 
potential and exact potentials strongly differ in this region and, at least for the Hooke's atom \cite{note3}, the i-LDA 
potential becomes much too confining at very large $r$'s.  However in this region $n^{\text{LDA}}(r)$ is still larger than the exact density.  Hence the interplay between many-body and single-particle self-interaction effects and between these effects and the non-locality of the mapping between density, $v_{\text{xc}}$ and external potential is highly nontrivial.  This is particularly evident for helium where the i-LDA potential is more attractive at the largest $r$ depicted.
Common to both systems is the drop in the i-LDA external potential after the crossing at intermediate $r$'s, this crossing is common to both potentials and densities.
We speculate that this feature is caused by a less investigated consequence of self-interaction, causing 
the electron density to spread out more at smaller $r$ in an attempt to 
minimize the self-Coulomb energy. As this starts to occur before the asymptotic
region is reached, it involves many electrons, and can thus be interpreted as
a consequence of the many-electron self-interaction error\cite{msie1,msie2}.

The self-interaction problem in a single-particle framework has been studied for many years. Much more recently, it has become apparent that this single-particle self-interaction correction 
does not fully remove the self-interaction error in many-particle systems. The search for a clear
indicator and a remedy for this problem has recently taken center stage as
one of the main unsolved problems of DFT\cite{msie1,msie2}.
Our reverse engineering prescription is a way to shed light on this complex
problem, since the many-electron self-interaction error contributes to the
difference between the true and the i-LDA potentials at intermediate distances.

These considerations illustrate the concept and use of the i-LDA system. 
To construct this system we have developed two iterative schemes for 
inverting the many-body Schr\"odinger equation. Previous inversion 
schemes within DFT either were concerned with inverting the much simpler 
single-particle KS equation, or directed at a one-dimensional model many-body 
system. Our schemes attain orders of magnitude higher accuracy than previous 
schemes and are applicable to three-dimensional many-body systems. 

Building on this technical advance, we have introduced the concept of 
{\em reverse engineering} in DFT and, more generally, in all many-body 
methods that introduce an effective single-body potential. This approach 
allows one to judge the performance and failures of, {\em e.g.}, an approximate 
density functional or single-particle equation, by comparing two external 
potentials, thus revealing spatially resolved information on properties such 
as self-interaction errors and non-locality, in a physically 
transparent way. 
Another possible application of quantum reverse engineering is to 
design that external potential that reproduces a desired density distribution 
in a given spatially inhomogeneous many-body system, which is an exciting 
prospect for the design of nanostructured devices.

JPC was supported by EPSRC-GB.  KC was supported by FAPESP and CNPq.  IDA and JPC acknowledge support from the Department of Physics of the University of York and the kind hospitality of IFSC-USP (Brazil).

\end{document}